\documentclass[pss]{wiley2sp} 
\usepackage{amsmath,amssymb}
\usepackage{bm}             

\begin{document}

\title{Reversible tuning the optical properties of defective TMDs monolayers}

\author{%
Mohammad Bahmani\textsuperscript{\Ast,\textsf{\bfseries 1}},
Michael Lorke\textsuperscript{\Ast,\textsf{\bfseries 1,2}},
Mahdi Faghihnasiri\textsuperscript{\textsf{\bfseries 3}},
Thomas Frauenheim\textsuperscript{\Ast,\textsf{\bfseries 1,4,5}}}

\mail{\textsf{mbahmani@uni-bremen.de}, \textsf{mlorke@itp.uni-bremen.de}, \textsf{frauenheim@bccms.uni-bremen.de}}

\institute{%
  \textsuperscript{1}\,Bremen Center for Computational Materials Science, Department of Physics, Bremen University, Am Fallturm 1, 28359 Bremen, Germany\\
  \textsuperscript{2}\,Institute for Theoretical Physics, Department of Physics, Bremen University, Otto-Hahn Alle, 28334 Bremen, Germany\\
  \textsuperscript{3}\,Computational Materials Science Laboratory, Nano Research and Training Center, 1478934371 Tehran, Iran
\\
  \textsuperscript{4}\,{Beijing Computational Science Research Center (CSRC), 100193 Beijing, China}
\\
  \textsuperscript{5}\,{Shenzhen JL Computational Science and Applied Research Institute, 518110 Shenzhen, China}}
  
%

             
\abstract{\bf%
Potential applications of monolayer of transition metal dichalcogenides (TMDs) in optoelectronic and flexible devices are under heavy investigation. Although TMDs monolayers are highly robust to external mechanical fields, their electronic structure is sensitive to compressive and tensile strain. Besides, intrinsic point defects are present in synthesized samples of these two dimensional (2D) materials which leads to the modification of their electronic and optical properties. Presence of vacancy complexes leads to absorption with larger dipole matrix elements in comparison to the case of simple transition metal vacancies. Using first principles calculations, we scrutinize the effect of various strain situations on the absorption spectra of such defective monolayers and show that strain engineering allows for reversible tuning of the optical properties.}

\maketitle 

%


\section{Introduction}
The semiconducting molybdenum~($Mo$) and tungsten~($W$) dichalcogenide monolayers (ML), with direct bandgap, high carrier mobility, and unique optical and mechanical properties, are widely investigated as they have shown promising applications in electronics and optoelectronics \cite{Mak2010,Wang2012,Mak2016,Lin2018}. Such 2D~TMDs exhibit superior photoluminescence (PL) in the visible range \cite{Wurstbauer2017,Mak2010}. It has been observed that in all the synthetic samples of MLs~TMDs, produced via the chemical vapor deposition and the mechanical exfoliation, structural imperfections are always present, in particular point vacancies \cite{Komsa2013,Zhou2013,Komsa2015,Lin2016}. At post growth phase, processes like vacuum annealing, ion bombardment, chemical treatment, or ion irradiation, can also be used to introduce such defects onto the monolyers \cite{Klein2018,Klein2019,Ghorbani-Asl2017}. As a consequence, the defect engineering has been proposed to tune the electronic and optical properties of 2D~TMDs \cite{VanDerZande2013,Butler2013,Ghorbani-Asl2013,Roldan2015,Khan2017}. Among these defects, vacancies induce localized defect states deep inside the band gap and close to the valance band maximum~(VBM) as well as the conduction band minimum~(CBM) \cite{Zhou2013,Haldar2015,Komsa2015,Pandey2016,Bahmani2020}. These midgap states lead to new optical transitions in the spectra, making the defective TMDs monolayers even more interesting for electronic and optoelectronic devices \cite{Yuan2014,Mak2016,Lin2016,Klein2019}. Particularly, defect luminescence centers are promising candidates for light-emitting diodes~(LEDs) and lasers \cite{Nan2014,Withers2015,Wu2015,Klein2019}. In addition, single-photon emission from defect levels~(DLs) inside the bandgap of monolayers of molybdenum disulfide~(MoS$_{2}$) and tungsten diselenide~(WSe$_{2}$) were observed \cite{Koperski2015,He2015,Tripathi2018,Klein2019}. It has been measured that photo-excited charge carriers can be trapped at the midgap localized levels, which, in turn, leads to a growth of the photocurrent in photodetectors based on MLs~MoS$_{2}$ \cite{Perea-Lopez2014,Amit2017,Ghimire2019}. Hence, scrutinizing the optical properties of crystalline point defects inside MLs~TMDs is of great importance both from the application and the fundamental point of view \cite{Lin2016,Mak2016,Wurstbauer2017,Lin2018}.

Experimental measurements have shown that MLs~TMDs are highly stable under external mechanical fields compared to conventional bulk semiconductors \cite{Bertolazzi2011,Castellanos-Gomez2012,Castellanos-Gomez2015,Lloyd2016}. At the same time, the electronic and optical properties of such 2D materials are proven to be sensitive to compressive and tensile strain \cite{Ghorbani-Asl2013,Roldan2015,Steinhoff2015,Lloyd2016,Frisenda2017}. For example, it was shown that MLs~TMDs undergo a direct-indirect bandgap transition via applying $1.5\%$ uniaxial strain \cite{Roldan2015,Ghorbani-Asl2013}. As a large biaxial strain is applied to MoS$_{2}$ monolayer, its bandgap can be reversibly and continuously tuned up to 500 meV \cite{Lloyd2016}. These unique features establish their great advantages over conventional semiconductors for applications in transparent and flexible electronic and optoelectronic devices \cite{Wang2012,Lopez-Sanchez2013,Withers2015,Lin2018}. Integrating onto microelectromechanical systems~(MEMS), strain of more than $1.3\%$ has been applied to MLs~MoS$_{2}$, which lays the ground for novel applications of 2D~TMDs in flexible LEDs and field-effect transistors~(FETs) \cite{Christopher2019}. Biaxial strain has been observed to tune the characteristics of photodetector devices based on MLs~MoS$_{2}$ \cite{Gant2019}. Recently, we showed the possibility to engineer the degenerate DLs of sulfur and molybdenum vacancies inside MLs~MoS$_{2}$ via various mechanical deformations \cite{Bahmani2020}. Here, the $C_{3v}$ symmetry of the hexagonal monolayers were broken by the applied strain, thus, splitting the degenerate DLs inside the bandgap up to 450 meV. 

In this paper, we study the effect of three different compressive and tensile strain on the optical properties of MoS$_{2}$ and WSe$_{2}$ monolayers, containing transition metal vacancies and vacancy complexes. Using density functional theory (DFT), we analyze the change in absorption spectra of such defective monolayers within the linear response regime. At~zero strain, the optical response from defects inside MLs~MoS$_{2}$ are different than vacancies inside MLs~WSe$_{2}$. Applying various mechanical deformations to MLs~TMDs modifies the absorption strength, depending on the type of vacancy. Several defect-defect transitions (DDTs) become visible if strain is applied to the defective monolayers. 

The paper is organized as follows: In section \ref{compdetails}, we explain the details of the computations method. In section \ref{resdis}, the main results of the paper are shown. We conclude with a summary and discussion in section \ref{con}.

\section{Computational Details}  \label{compdetails}
Semiconducting TMDs monolayers, with hexagonal symmetry, are constructed of a triple X-M-X layer, where the transition metal M is covalently bonded to two chalcogen atoms X \cite{Wang2012,Roldan2015,Lin2016}. In the present work, we focus on two polytypes: MoS$_{2}$ and WSe$_{2}$, which are also the experimentally most investigated ones. First-principles calculations are performed using the DFT formalism as implemented in the SIESTA code \cite{Ordejon1996,Soler2002}. The wavefunction for the valance electrons are expanded by a linear combination of double-zeta basis sets with polarization function (DZP). The 4p diffusive orbitals are also included to improve the characterization of sulfur atoms. The exchange and correlation interactions are described using the generalized gradient approximation (GGA) via the semilocal XC-functional of Perdew-Burke-Ernzerhof (PBE) \cite{Perdew1996}. The GGA-PBE functional underestimates the bandgap, in particular for the case of TMDs monolayers. However, previous studies have shown that using more sophisticated methods would only lead to similar relative shifts in the band edges and DLs, but do not change the qualitative picture of the defect states within the bandgap \cite{Refaely-Abramson2018,Naik2018,Schuler2019}.
Within the Troullier-Martin approach, we have generated a norm-conserving and relativistic pseudopotentials, including core electrons, to describe the valance electrons \cite{Troullier1991A,Troullier1991B}. Values for the Energy-Shift and the SplitNorm are equal to $0.02 Ry$ and $0.16$, respectively. Lattice vectors and atomic positions of the equilibrium and strained configurations are optimized using the conjugate-gradient (CG) method. The lower limit for the Hellmann-Feynman forces on each atom is set to 0.01~$eV/\AA$. In order to minimize spurious defect-defect interactions between the defect images in adjacent supercells as much as possible, we calculate the properties of defective MoS$_{2}$ (WSe$_{2}$) monolayers for supercell sizes of 6$\times$6$\times$1 to 9$\times$9$\times$1. Our intention is to investigate the properties of isolated defects, rather than the influence of defect concentration. Accordingly, we found that for practical calculations, monolayers of 8$\times$8$\times$1 are a good compromise between accuracy and numerical efforts, for transition metal vacancies as well as for vacancy complexes. 
On the other hand, this supercell sizes are too small to observe ripple structures due to compressive strain. 
Normal to the layers, a vacuum of 40~$\AA$ is considered to avoid interactions between adjacent monolayers. The convergency of the total energy is ensured as the difference between two consecutive self-consistent field steps is set to less than $10^{-4}$~$eV$. The Brillouin zone of supercells is sampled by a 5$\times$5$\times$1 k-points in the Monkhorst-Pack scheme to obtain both geometries and electronic properties. The Hartree, exchange, and correlation contribution to the total energy in the real space are calculated using a mesh cut-off of 450~$Ry$. The energy cut-off and k-points are considered converged when total energy differences were below $10^{-4}$~$eV$ and $10^{-5}$~$eV$, respectively. According to previous theoretical studies, the ground state of defective MLs~TMDs is non-magnetic up to more than $5\%$ strain \cite{Tao2014,Zheng2014,Yun2015,Li2018a}. Since we only applied $2\%$ of mechanical deformations to the monolayers, spin-polarization is not considered in this paper. The qualitative picture of the electronic structure of TMDs monolayers containing point defects is preserved in the presence of the spin-orbit coupling (SOC), even though SOC splits the VBM \cite{Refaely-Abramson2018,Naik2018,Schuler2019}. Due to the fact that it has minor influence on the provided analysis and the final conclusions, SOC is neglected in the present work.

\begin{figure}[h]
\begin{center}
\includegraphics[width=0.5\textwidth]{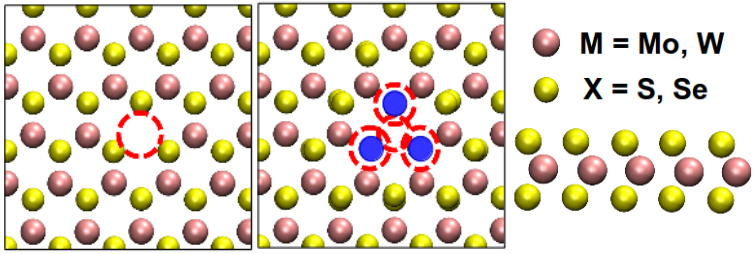}
\caption{(Color online) Left) A monolayer of MoS$_{2}$ (WSe$_{2}$) containing the point defects: $V_{M}$ and $V_{M+3X}$. In every calculation, we only focus on one of these vacancies inside the monolayers. Red dashed-circles denote the position of the missing metal or chalcogens inside the monolayers. Right) side-view of MLs~TMDs. Here, $M = Mo,W$, and $X = S,Se$.}
\label{fig:1}
\end{center}
\end{figure}
Fig.~\ref{fig:1} shows the position of the point vacancies and their neighboring atoms inside MLs~MoS$_{2}$ (MLs~WSe$_{2}$). In this paper, we study MoS$_{2}$ and WSe$_{2}$ monolayers with a transition metal vacancy $V_{M}$ and vacancy complexes $V_{M+3X}$, where $M$ is $Mo$,$W$, and $X$ is $S,Se$. There are computational studies in addition to experimental observation of such vacancies in MLs~TMDs samples \cite{Zhou2013,Komsa2015,Lin2016,Ma2017,Stanford2019}. It is as well possible to introduce these vacancies by plasma exposure and ion-irradiation \cite{Ghorbani-Asl2017,Klein2019,Stanford2019}. For both defect cases, the $C_{3v}$ symmetry of hexagonal structures are preserved, as shown in Fig.~\ref{fig:1}. The atomic geometries are depicted via the VMD tool \cite{HUMP96}.

\begin{figure}[!htb]
\begin{center}
\includegraphics[width=0.5\textwidth]{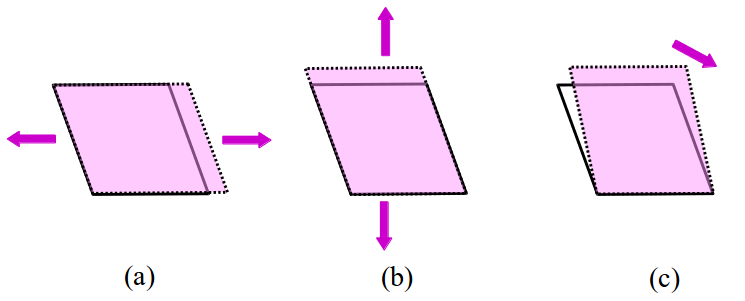}
\caption{(Color online) schematic view of the uniaxial strain in a) X-direction, and b) Y-direction, c) sheer T1 strain.}
\label{fig:2}
\end{center}
\end{figure}

We examine the effect of three different compressive and tensile strain on the optical properties of defective MoS$_{2}$ (WSe$_{2}$) monolayers, which are shown in Fig.~\ref{fig:2}. For the purpose of resembling simple deformations, we consider uniaxial strain in X- and Y-direction. There is also an inhomogeneous shear type strain (shear T1) which maintains the magnitude of in-plane lattice vectors but changes the angle between them. These types of strain are calculated at $2\%$ of compression and $2\%$ of stretching, yet they are below the breaking point of the monolayers \cite{Bertolazzi2011}.

Optical properties are calculated using the SIESTA code, in which, the linear response function is used to compute the imaginary part of the dielectric function $\text{Im}[\varepsilon(\omega)]$ \cite{Yu2013}: 
\begin{equation}
\begin{aligned}
Im[\varepsilon(\omega)] = & \frac{1}{4\pi \varepsilon_0} \left( \frac{2\pi e}{m \omega}\right) ^2 \sum_{\textbf{k}} \vert \textbf{p}_{c,v}\vert^2 \\
& \times \delta (E_c(\textbf{k}) - E_v(\textbf{k}) - \hbar\omega) \\
& \times \left[ f(E_v(\textbf{k})) - f(E_c(\textbf{k})) \right].
\end{aligned}
\end{equation}
Here, \textit{c} and \textit{v} subscripts are denoting conduction and valance bands properties, respectively. $E_{c,v}(\textbf{k})$ are the energy bands with \textit{k}-vector \textbf{k}. Parameter \textit{m} is the electron mass, \textit{$\hbar\omega$} is the photon energy, and \textit{$\textbf{p}_{c,v}$} is the momentum operator. An optical mesh size and broadening of 25$\times$25$\times$1 and 0.02 $eV$ are chosen, respectively.

We focus here on the dipole transition strength of various inter DL transitions under the influence of different types of strain. Although including the many-body effects modify the electronic and optical properties of TMDs monolayers, there are strong experimental evidence, as in Refs. \cite{Klein2018,Klein2019,Barthelmi2020,Mitterreiter2020}, of peaks at energies below the bandgap. These peaks were identified to be corresponding to DDTs. Considering such effects into the calculations does not change the characteristics of optically active and inactive states. Besides, in order to avoid defect-defect interactions, we study a very large system size, for which it is prohibitive to include a description of the level of many-body perturbation theory (like a GW0+BSE approach).
Thus, we have neglected the electron-hole Coulomb interactions in this work. 
As we consider only linear optical absorption and hence probing the states which would be available for excitation, geometry relaxation via the excitation process does not play a role here, in contrast to studies based on photoluminescence, where actual electrons are excited.

\section{Results and Discussion} \label{resdis}
We investigate the effect of various strain situations on the optical properties of MoS$_{2}$ and WSe$_{2}$ monolayers containing transition metal vacancies, $V_{M}$, as well as vacancy complexes, $V_{M+3X}$. For this purpose, the optical spectra are investigated via the imaginary part of the dielectric function, $\text{Im}[\varepsilon]$. The presence of such vacancies in the synthesized samples have been observed in atomic-resolution measurements and studied via theoretical methods \cite{Zhou2013,Komsa2015,Ma2017,Stanford2019}. Moreover, these defects can also be introduced to the TMDs monolayers by post processing techniques, such as plasma exposure and ion-irradiation \cite{Ghorbani-Asl2017,Klein2019,Stanford2019}.

\subsection{Electronic structure}
\begin{figure}[!htb]
\begin{center}
\includegraphics[width=0.40\textwidth]{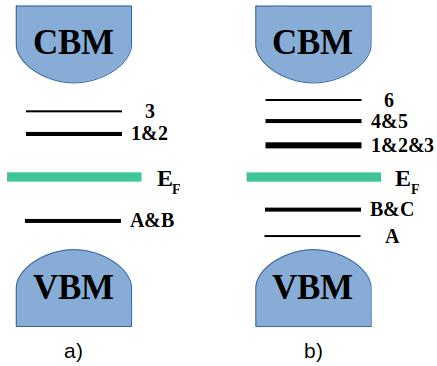}
\caption{(Color online) Schematic of the electronic structure of the defective MLs~MoS$_2$~(MLs~WSe$_2$) with a) $V_{Mo}$ ($V_{W}$), b) $V_{Mo+3S}$ ($V_{W+3Se}$). Valance and conduction band edges, VBM and CBM, are shown along with the Fermi energy~($E_F$). Occupied (unoccupied) DLs in the bandgap are labeled with letters (numbers). A\&B,~B\&C,~1\&2,~4\&5 are double-degenerate states. 1\&2\&3 are triple-degenerate DLs.}
\label{fig:3}
\end{center}
\end{figure}

A schematic of the electronic structure of these defective monolayers is displayed in Fig.~\ref{fig:3}. Here, we show the VBM, the CBM, and the Fermi energy ($E_F$) as well as the DLs. The occupied DLs are named with letters A to C, while the unoccupied DLs are labeled with numbers 1~to~6. 
Based on first-principles DFT analysis, there are five DLs in the band structure of MoS$_2$ (WSe$_2$) monolayers with a single transition metal vacancy, $V_{Mo}$ ($V_{W}$), as shown in Fig.~\ref{fig:3}a. Two occupied double-degenerate levels are labeled A\&B, while 1\&2 are the double-degenerate unoccupied DLs and $3$ is a non-degenerate state. Structures with vacancy complexes of $V_{Mo+3S}$ and $V_{W+3Se}$ are shown in Fig.~\ref{fig:1}. Figure~\ref{fig:3}b shows a schematic of their electronic structure; 
 B\&C are double-degenerate occupied DLs. There are also six unoccupied localized states, a triple-degenerate 1\&2\&3, a double-degenerate 4\&5, and a non-degenerate level 6. These results are in accordance with previous reports \cite{Zhou2013,Komsa2015,Haldar2015,Pandey2016,Bahmani2020}.  

Accordingly, DDTs are indicated corresponding to the levels involved in the transitions: A1,A2,\ldots,B1,\ldots,C6. The orbital characteristics of the localized states and VBM are scrutinized to identify the DDTs in the absorption spectra of such monolayers. Major contributions are originated from the $d$ orbitals of transition metals ($Mo, W$) and $p$ orbitals of chalcogens ($S,Se$). In the case of simple metal vacancy in MLs~MoS$_{2}$, VBM is built from $4d_{xy}$,$4d_{x^2-y^2}$~($4d_{XY}$) and $3p_x$,$3p_y$ orbitals, while a mixture of $4d_{z^2}$,$4d_{XY}$ and $3p_z$ orbitals are mainly involved in the VBM of system with $V_{Mo+3S}$. For $V_{W}$ in MLs~WSe$_{2}$, $5d_{z^2}$ and $5d_{xy}$,$5d_{x^2-y^2}$~($5d_{XY}$) orbitals are hybridized with $4p_x$,$4p_y$,$4p_z$ ($4p$) orbitals to construct the VBM; however, only $5d_{z^2}$,$5d_{XY}$ orbitals contribute to the VBM when $V_{W+3Se}$ is present in monolayers. These findings are in line with previous studies \cite{Cappelluti2013,Gonzalez2016,Pandey2016}.
Orbital characteristics of the DLs for each vacancy are presented in corresponding sections.

\subsection{Optical properties}
We investigate the effect of various types of strain on the absorption spectra of these defective monolayers within the linear response regime. Peaks at energies below the bandgap indicate the presence of optically active DLs within the electronic structure of defective monolayers.

\subsection*{Simple transition metal vacancy}
\begin{figure*}[!htb]
\begin{center}
\includegraphics[width=1.00\textwidth]{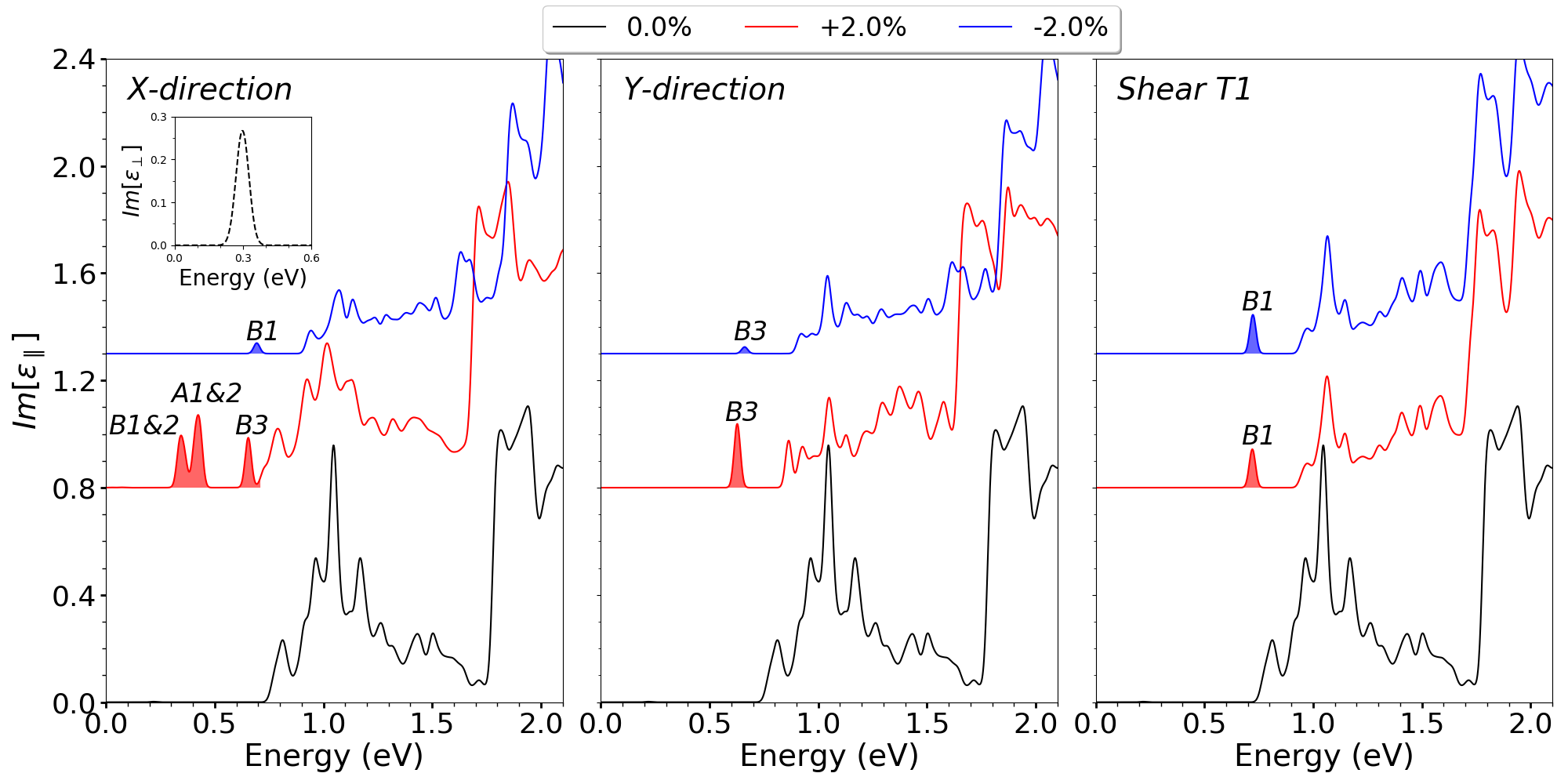}
\caption{(Color online) The absorption spectra for MLs~MoS$_{2}$ with $V_{Mo}$ at zero strain (black lines) and under $2\%$ compressive (blue lines) and $2\%$ tensile (red lines) strain. The inset is the out-of-plane response of the defective monolayer. From left to right, the plots represent the effect of strain in X-direction, Y-direction, and shear T1 strain. For each strain case, DDTs are labeled and highlighted with shaded colors.}
\label{fig:4}
\end{center}
\end{figure*}

\hspace{0.1cm} The optical properties of MLs~MoS$_{2}$ with $V_{Mo}$ vacancy are shown in Fig.~\ref{fig:4}. We study the in-plane imaginary part of the dielectric function ($\text{Im}[\varepsilon_{\parallel}(\omega)]$) for the unstrained case (black lines) and under various types of compressive (blue lines) and tensile (red lines) strain. 
Orbitals $4d_{xz}$,$4d_{yz}$ and $3p_x$,$3p_y$,$3p_z$~($3p$) are main components of two occupied states A and B, while unoccupied DLs are constructed from a mixture of $4d_{z^2}$,$4d_{XY}$ and $3p$ orbitals. Hence, corresponding elements of the dipole matrix have zero value, leading to optically inactive DDTs in the in-plane spectra (See black curves in Fig.~\ref{fig:4}). Two peaks with the lowest energy arise due to the transition between the VBM and unoccupied double-degenerate DLs. In the inset of Fig.~\ref{fig:4}, the out-of-plane absorption spectra ($\text{Im}[\varepsilon_{\perp}(\omega)]$) is plotted for unstrained defective monolayers. In contrast to the case of in-plane spectra, DDTs A1\&2 and B1\&2 are visible here. These states are mainly composed of surrounding atomic orbitals with components outside of the XY-plane. Fig.~\ref{fig:4} also shows the effect of various strain situations on the spectra of $V_{Mo}$ in MLs~MoS$_{2}$, where DDTs are indicated with shaded colors. As uniaxial and inhomogeneous shear T1 strain are applied, transitions between occupied and unoccupied DLs become optically active, due to the change in the hybridization of orbitals surrounding the vacancy. This stems from breaking the hexagonal symmetry of the defective MLs~MoS$_{2}$ via strain, thus removing the degeneracy of the localized states \cite{Ahn2017,Sensoy2017,Bahmani2020}. It can also be explained via the modification of the geometry as function of applied strain, which is elaborated on in the Appendix. In the case of $2\%$ stretching in X-direction, three peaks can be observed corresponding to DDTs B1\&2, A1\&2, and B3. Orbital characteristics of occupied level A~(B) is modified and now containing a mixture of $4d_{xy}$ and $3p$ ($4d_{x^2-y^2}$,$4d_{z^2}$ and $3p$) orbitals. The first two unoccupied DLs remain intact under this strain, however, the third unoccupied localized state is now mainly constructed of $4d_{z^2}$,$4d_{XY}$ and $3p$ orbitals. Modifications of the electronic structure under $2\%$ of compressive strain leads to observation of B1 transition, where both states are originated from a combination of $4d_{z^2}$,$4d_{XY}$ and $3p$ orbitals. When uniaxial compressive or tensile strain in Y-direction is applied, B3 transition becomes visible in the in-plane response of the dielectric function, where mostly $4d_{z^2}$,$4d_{XY}$ and $3p$ orbitals are contributing to both DLs. In general, the absorption strength of DDTs is larger for uniaxial tensile than compressive strain. The influence of compressive and tensile shear T1 strain on the optical response of MoS$_{2}$ monolayers with $V_{Mo}$ are similar and lead to a peak stemming from B1 transition. Here, $4d_{z^2}$,$4d_{XY}$ and $3p$ orbitals are mainly contributing to these states. 

\begin{figure*}[!htb]
\begin{center}
\includegraphics[width=1.00\textwidth]{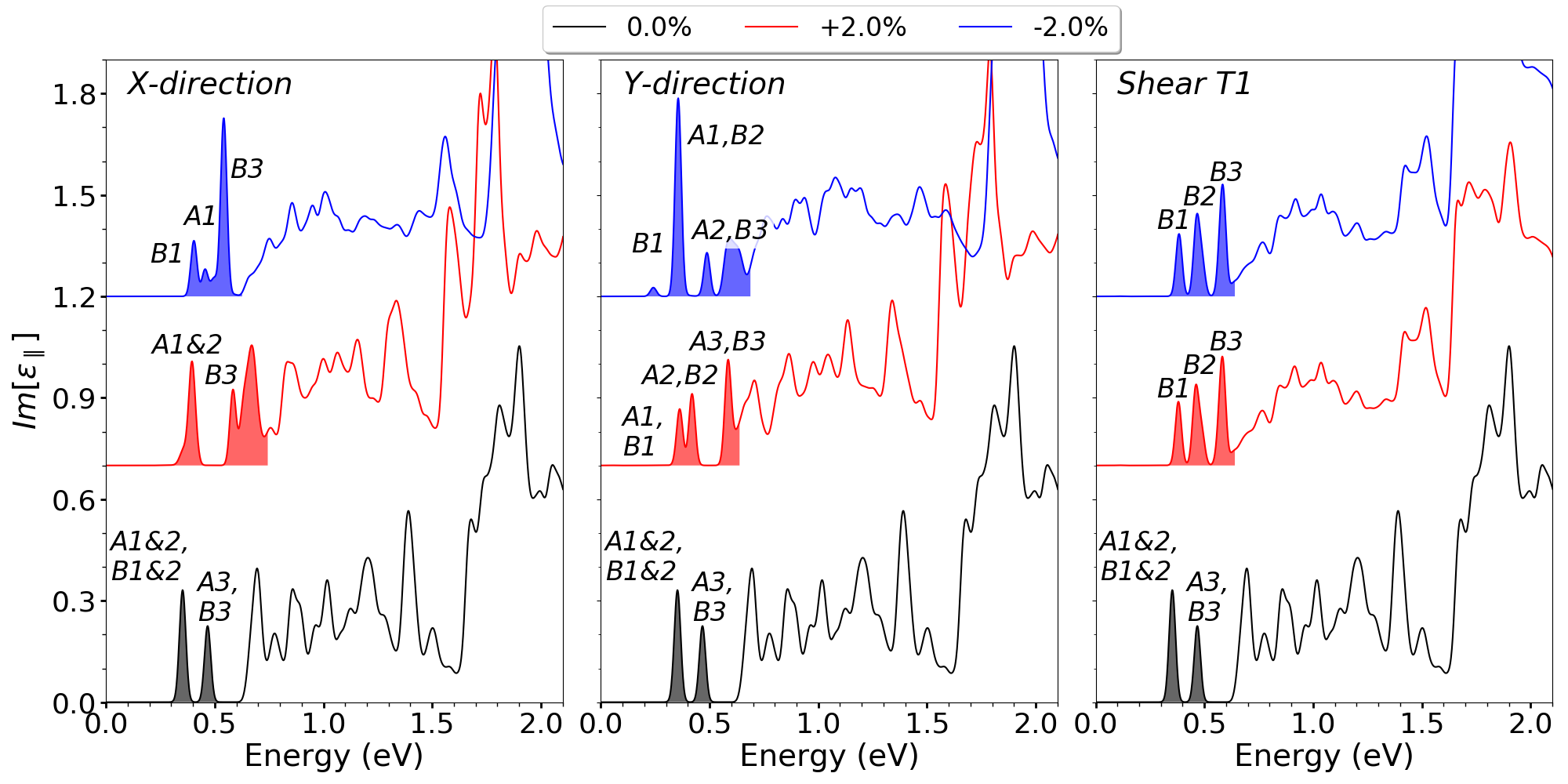}
\caption{(Color online) The absorption spectra for MLs~WSe$_{2}$ with $V_{W}$ at zero strain (black lines) and under $2\%$ compressive (blue lines) and $2\%$ tensile (red lines) strain.  From left to right, the plots represent the effect of strain in X-direction, Y-direction, and shear T1 strain. For each strain case, DDTs are labeled and highlighted with shaded colors.}
\label{fig:5}
\end{center}
\end{figure*}

The presence of a simple metal vacancy, $V_{W}$, in MLs~WSe$_{2}$ results in localized states inside the bandgap. In Fig.~\ref{fig:5}, the optical spectra are shown (black lines) for the \,unstrained defective structure. 
Two peaks corresponding to transitions A1\&2,B1\&2 and A3,B3 are highlighted. The fact that DDTs are visible in the spectra is in explicit contrast to the case of $Mo$ vacancy in MLs~MoS$_{2}$ at zero strain. This translates into differences in orbital characteristics of the DLs in two materials. For the case of $V_{W}$ in MLs~WSe$_{2}$, two occupied localized states A and B are constructed of $5d$ orbitals mixing with $4p$ \,orbitals, and three unoccupied DLs have major contributions from $5d_{z^2}$,$5d_{XY}$ and $4p$ orbitals. On the other hand, localized states in unstrained MLs~MoS$_{2}$ with $V_{Mo}$ are mainly constructed from $4d_{xz}$,$4d_{yz}$ and $3p$ \,orbitals. Comparing Figs.~\ref{fig:4} and~\ref{fig:5}, it can be seen that dipole matrix elements are larger for $V_{W}$ than $V_{Mo}$. In Fig.~\ref{fig:5}, the effect of various strain situations on the absorption spectra of these defective WSe$_{2}$ monolayers are displayed where DDTs are labeled and shown with shaded colors. In the case of $2\%$ uniaxial tensile strain in X-direction, although the orbital characteristics of the unoccupied states remain unchanged, states A and B are now dominated with $5d_{xz}$,$5_{dxy}$,$5d_{yz}$,$5d_{z^2}$,$4p$ and $5d_{z^2}$,$5d_{XY}$,$4p$ orbitals, respectively. Peaks corresponding to DDTs A1\&2 and B3 are visible. Applying $2\%$ of compressive strain in X-direction, three transitions B1, A1, and B3 can be observed. Here, there are modifications only in the major orbitals contributing to the states A and B as well as the first unoccupied level. 
When uniaxial strain in Y-direction are applied, hybridization of the occupied DLs A and B are changed, while the main orbital components of the unoccupied states stay unaffected. Thus, the degeneracy breaking leads to the observation of A1--3 and B1--3 transitions in the spectra. When $2\%$ \,compressive or tensile shear T1 strain is applied to MLs~WSe$_{2}$ with $V_{W}$, orbitals $5d_{z^2}$,$5d_{XY}$ and $4p$ have the largest coefficients in wavefunction expansion of states A and B. The first unoccupied DL is now constructed of $5d$ and $4p$ orbitals. Peaks corresponding to DDTs B1, B2, and B3 are shown in Fig.~\ref{fig:5}.

\subsection*{Vacancy complexes}
\hspace{0.1cm} In synthesized samples as~well~as during post processing mechanisms, vacancy complexes, i.e. $V_{Mo+3S}$ and $V_{W+3Se}$, are more likely to be present than single transition metal vacancies, according to previous studies \cite{Komsa2015,Gonzalez2016,Sensoy2017,Ahn2017,Bahmani2020}. This is particularly important when generating vacancies for single-photon emitters at selective sites \cite{Ghorbani-Asl2017,Klein2019}. We investigate the optical properties of MoS$_{2}$ and WSe$_{2}$ monolayers containing vacancy complexes at zero strain and under various mechanical deformations. Such studies are also of interest for the production of flexible optoelectronic devices. In comparison to the case of $V_{Mo}$ ($V_{W}$) vacancies in MLs~MoS$_{2}$ (MLs~WSe$_{2}$), out-of-plane optical responses are negligible for monolayers with the complex vacancies, $V_{Mo+3S}$ ($V_{W+3Se}$).

\begin{figure*}[!htb]
\begin{center}
\includegraphics[width=1.02\textwidth]{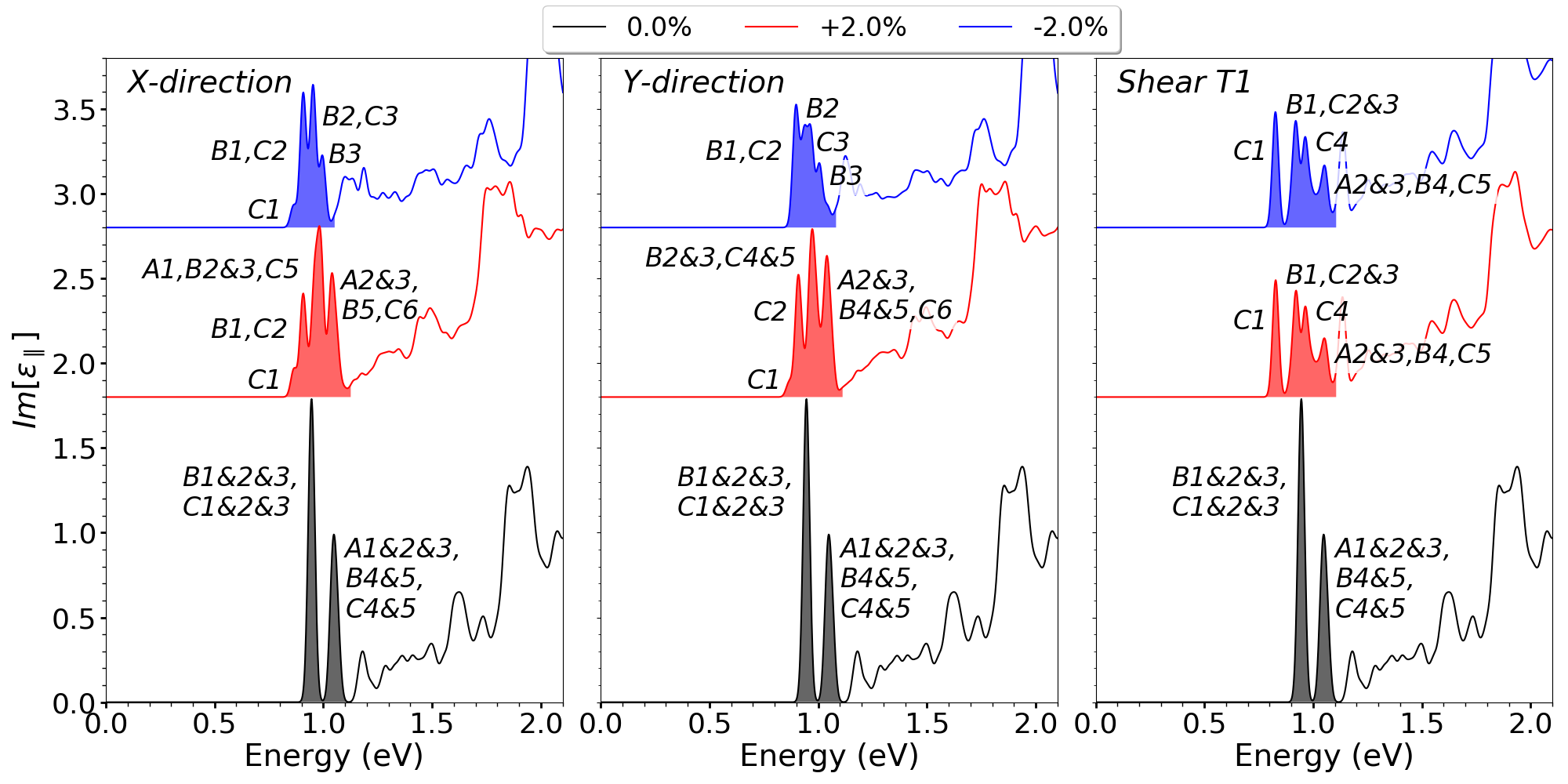}
\caption{(Color online) The absorption spectra for MLs~MoS$_{2}$ with $V_{Mo+3S}$ at zero strain (black lines) and under $2\%$ compressive (blue lines) and $2\%$ tensile (red lines) strain. From left to right, the plots represent the effect of strain in X-direction, Y-direction, and shear T1 strain. For each strain case, DDTs are highlighted with shaded colors and their labels.}
\label{fig:6}
\end{center}
\end{figure*}

Shown in Fig.~\ref{fig:6}, the in-plane imaginary part of the dielectric function is plotted for MoS$_{2}$ monolayers with $V_{Mo+3S}$. 
In this figure, optically active DDTs are highlighted. Their dipole matrix elements are noticeably much larger than peaks in the absorption spectra of MLs~MoS$_{2}$ with $V_{Mo}$ (See Fig.~\ref{fig:4}). This could be an evidence that in ion-irradiation processes, single photons are emitted from vacancy complexes rather than simple transition metal vacancies. The peak with the highest absorption comes from DDTs B1\&2\&3, and C1\&2\&3, i.e. transitions from double-degenerate DLs B\&C to triple-degenerate unoccupied localized states 1\&2\&3. The other peak is originated from transitions A1\&2\&3, B4\&5, and C4\&5. Unoccupied DLs 4\&5 are double-degenerate. At zero strain, all the DLs are mainly composed of $4d$ and $3p$ orbitals. Even though the degeneracy of DLs is broken via applying uniaxial and inhomogeneous strain, orbital contributions to the localized states remain untouched. Except for the case of tensile strain in X- and Y-direction, where the occupied state A is mostly constructed from a mixture of $4d$ and $3p_{y}$,$3p_{z}$ orbitals. As monolayers are stretched or compressed by $2\%$ of uniaxial strain, four optically active DDTs can be seen in the spectra. These are indicated in Fig.~\ref{fig:6} with their corresponding labels. \,Applying $2\%$ of either compressive or tensile shear T1 strain leads to identical modifications in the optical spectra, i.e. four peaks corresponding to transitions C1, B1,C2\&3, C4, and A2\&3,B4,C5 can be observed. All being said, the absorption strength of DDTs, consequently their brightness, is reduced by a factor of almost two (three) via uniaxial (inhomogeneous shear T1) strain. 

\begin{figure*}[!htb]
\begin{center}
\includegraphics[width=1.02\textwidth]{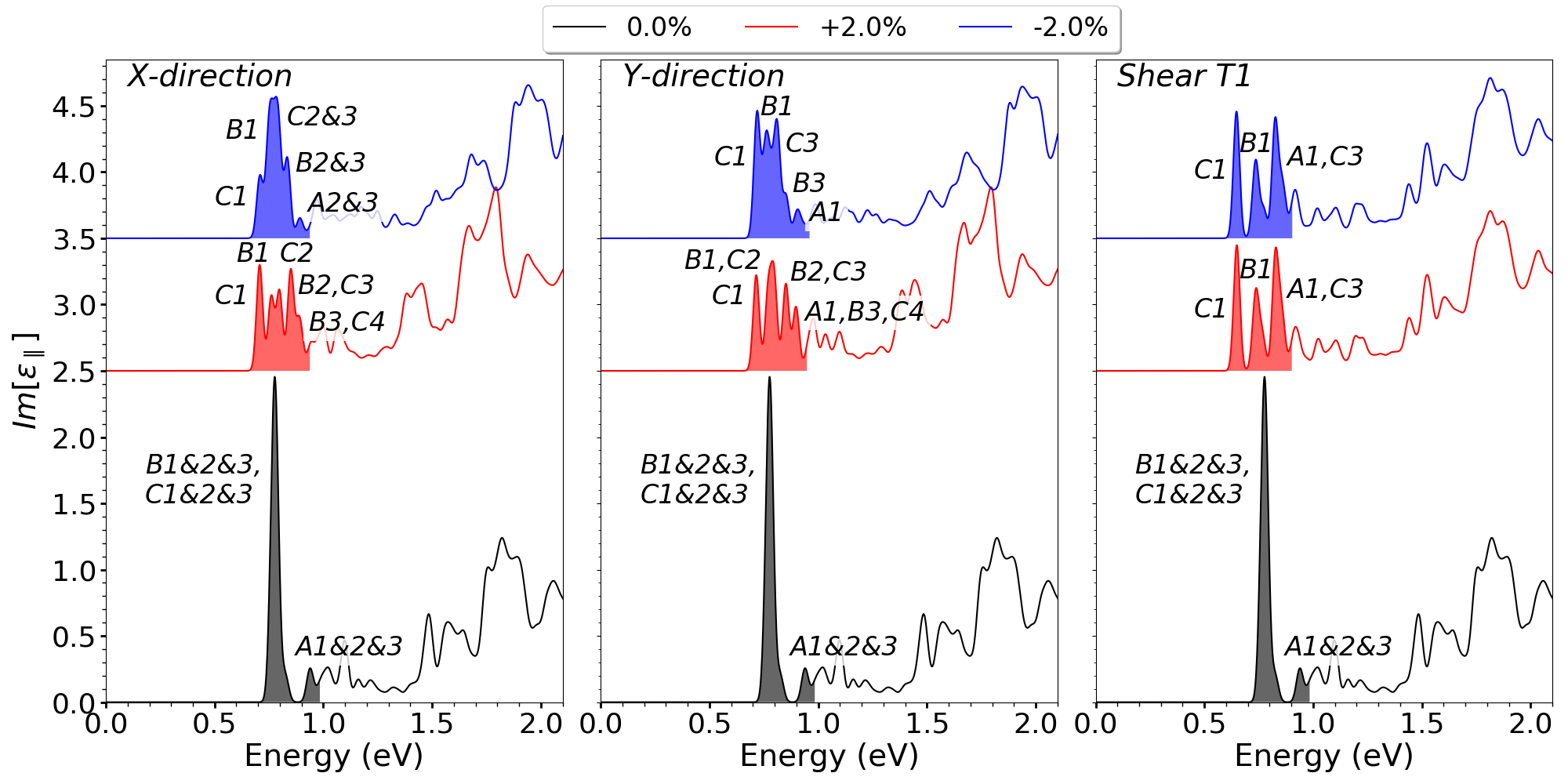}
\caption{(Color online) The absorption spectra for MLs~WSe$_{2}$ with $V_{W+3Se}$ at zero strain (black lines) and under $2\%$ compressive (blue lines) and $2\%$ tensile (red lines) strain. From left to right, the plots represent the effect of strain in X-direction, Y-direction, and shear T1 strain. For each strain case, DDTs are highlighted with shaded colors and their labels.}
\label{fig:7}
\end{center}
\end{figure*}

The absorption spectra of unstrained MLs~WSe$_{2}$ with the vacancy complex $V_{W+3Se}$ is shown with black curves in Fig.~\ref{fig:7}. 
Two sets of optically active DDTs can be seen, i.e. B1\&2\&3,C1\&2\&3 and A1\&2\&3. It should be noted that B\&C are double-degenerate occupied states, and 1\&2\&3 are triple-degenerate unoccupied levels. Interestingly, the dipole matrix elements of the first peak is about five times larger than the intensity of the peaks in the optical response of MLs~WSe$_{2}$ with $V_{W}$ (See Fig.~\ref{fig:5}). This outcome emphasizes our hypothesis that in ion-irradiated samples, the main source of single-photon emissions could be vacancy complexes rather than single transition metal vacancies. At zero strain, $5d$ orbitals of $W$ have the largest coefficients in wavefunction expansion of state A, which stays untouched as all strain situations are applied. Orbital characteristics of the occupied DLs B and C are mainly a combination of $5d$ and $4p$ orbitals. 
In the case of the unstrained defective monolayers, major contributions to all unoccupied DLs come from $5d$ and $4p$ orbitals which remain the same under any types of strain. In Fig.~\ref{fig:7}, blue (red) lines show the impact of various compressive (tensile) strain on the optical spectra of the defective monolayers, where DDTs are indicated with shaded colors and corresponding labels. When $2\%$ of compression or tensile uniaxial strain is applied to the defective MLs~WSe$_{2}$, degeneracy of the occupied DLs B and C is removed due to changes in their orbital characteristics. This leads to the observation of several optically active DDTs in the spectra, as indicated in Fig.~\ref{fig:7}. Shear T1 compressive and tensile strain result in analogous modifications to the absorption spectra of MLs~WSe$_{2}$ with $V_{W+3Se}$. Three peaks corresponding to DDTs C1, B1 and A1,C3 can be seen in the spectra. Here, only the orbital components of the occupied state B is changed to a mixture of $5d$ and $4p_x$,$4p_y$ orbitals. It can be observed in Fig.~\ref{fig:7} that the absorption strength are decreased by a factor of almost three as uniaxial or inhomogeneous strain are applied, for both compressing and stretching. 

\section{Conclusion}   \label{con}
In this paper, we have investigated the optical properties of MLs~MoS$_{2}$ and MLs~WSe$_{2}$ containing point vacancies; $V_{M}$ and $V_{M+3X}$. The optical spectra are calculated using DFT. At zero strain, it is shown that DDTs are visible in the in-plane spectra for $V_{W}$ in MLs~WSe$_{2}$ in contrast to MLs~MoS$_{2}$ with $V_{Mo}$. According to our study, dipole matrix elements of peaks originated from DDTs are significantly larger for the case of the vacancy complexes than the simple vacancies. 

The effect of two uniaxial and an inhomogeneous shear T1 strain on the optical properties of defective TMDs monolayers is studied. Interestingly, DDTs in MLs~MoS$_{2}$ with $V_{Mo}$ become visible in the in-plane spectra as mechanical deformations are applied. We trace this behavior back to the change in the hybridization of atomic orbitals surrounding the vacancy at the defect site. Depending on the type of strain, the absorption strength of MLs~TMDs with vacancy complexes has been reduced by a factor of two to three. Thus, the brightness of the spectra from samples with point vacancies could be reduced via strain. Applying strain allows to tune the optical properties of monolayers in a controllable way. Our findings will be beneficial to the application of MLs~MoS$_{2}$ in optoelectronic, flexible, and piezoelectric devices as well as heterostructure setups.

\begin{acknowledgement}
We thank the DFG funded research training group "GRK2247". M.B. acknowledges the support provided by DAAD and the PIP program at Bremen university. M.B. also thanks Dr. Miguel Pruneda for his help to produce well-performed pseudopotentials and basis sets.
\end{acknowledgement}
\bibliographystyle{pss}
\bibliography{strain-defect-optic}

\appendix
\setcounter{figure}{0}
\renewcommand\thefigure{A.\arabic{figure}} 
\section*{Appendix}  \label{Appendix}
\begin{figure}[!htb]
\begin{center}
\includegraphics[width=0.45\textwidth]{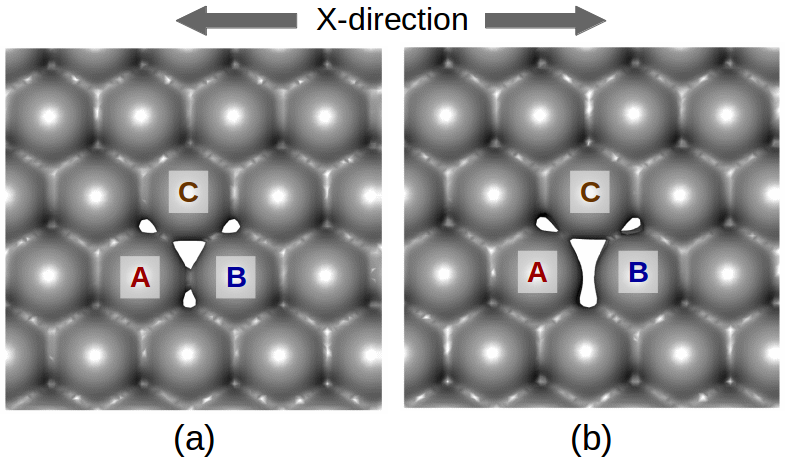}
\caption{(Color online) The change in the charge density of the MoS$_2$ monolayers containing $V_{Mo}$ under strain in X-direction for an amount of a) $0.0\%$ and b) $+2.0\%$. These are plotted at 0.095 $e/\AA^{3}$. Atoms around the vacancy are labeled with \textit{A}, \textit{B}, and \textit{C}.}
\label{fig:app_charg_den}
\end{center}
\end{figure}

\hspace{0.1cm} In this appendix, we highlight the relation between the change in the geometrical and electronic properties of defective MLs~MoS$_2$ and the modification of their optical response. In Fig. \ref{fig:app_charg_den}, the charge density of the monolayers with $V_{Mo}$ are shown for (a) $0.0\%$ and b) $+2.0\%$ strain in X-direction at isovalue of 0.095 $e/\AA^{3}$. Atoms around the vacancy are labeled with \textit{A}, \textit{B}, and \textit{C}. At zero strain, symmetric structure of the monolayer is untouched and the distance between the neighboring atoms is $3.238\AA$ for MLs~MoS$_2$ with V$_{Mo}$ and $3.176\AA$ for pristine geometry. When $2.0\%$ of tensile strain in X-direction is applied, the \textit{AC} and \textit{BC} distances are $3.262\AA$ and $3.192\AA$ for defective monolayer and pristine structure, respectively. However, since atoms at \textit{A} and \textit{B} positions are along the strain direction, the \textit{AB} distance for the case Mo vacancy in MLs~MoS$_2$ has been increased by $5.8\%$ to $3.426\AA$, which is much larger than for pristine structure with the expected $2.0\%$ increase to $3.240\AA$. This geometry modification is responsible for the changes in optical properties for such defective monolayers, as shown in Fig. \ref{fig:4}.

\end{document}